\documentclass{aa}

\usepackage{amsmath,amsfonts,amssymb}
\usepackage{bm}
\usepackage[table,  dvipsnames]{xcolor}
\usepackage{hyperref}
\usepackage{nicefrac}
\usepackage{placeins}
\usepackage{microtype}

\usepackage[varg]{txfonts}
\usepackage[T1]{fontenc}

\usepackage{soul}
\usepackage[normalem]{ulem}

\hypersetup{
	pdftitle = {Planet-driven gap opening in radiative disks: getting three dimensions stuffed into two},
	pdfauthor = {A. Ziampras, A. J. Cordwell, R. R. Rafikov, R. P. Nelson},
	pdfsubject = {},
	colorlinks = {true},
	linkcolor = [rgb]{0,0.35,0.7},
	citecolor = [rgb]{0,0.35,0.7},
	filecolor = [rgb]{0.61,0,0},
	urlcolor = [rgb]{0.61,0,0},
}

\usepackage{graphicx}

\newcommand{\DP}[2]{\frac{\partial{#1}}{\partial{#2}}}
\newcommand{\D}[2]{\frac{\text{d}{#1}}{\text{d}{#2}}}

\newcommand{\G}{\text{G}}
\newcommand{\Mstar}{M_\star}
\newcommand{\Lstar}{L_\star}
\newcommand{\Rp}{R_\mathrm{p}}
\newcommand{\Mp}{M_\mathrm{p}}
\newcommand{\hp}{h_\mathrm{p}}
\newcommand{\Hp}{H_\mathrm{p}}

\newcommand{\Mth}{M_\mathrm{th}}

\newcommand{\Msun}{\mathrm{M}_\odot}
\newcommand{\Lsun}{\mathrm{L}_\odot}

\newcommand{\Mearth}{\mathrm{M}_\oplus}
\newcommand{\Rgas}{\mathcal{R}}
\newcommand{\cs}{c_\mathrm{s}}

\newcommand{\OmegaK}{\Omega_\mathrm{K}}

\newcommand{\tauR}{\tau_\mathrm{R}}
\newcommand{\tauP}{\tau_\mathrm{P}}

\newcommand{\taueff}{\tau_\mathrm{eff}}
\newcommand{\kappaR}{\kappa_\mathrm{R}}
\newcommand{\kappaP}{\kappa_\mathrm{P}}
\newcommand{\cv}{c_\mathrm{v}}
\newcommand{\rhomid}{\rho_\mathrm{mid}}
\newcommand{\sigmaSB}{\sigma_\mathrm{SB}}
\newcommand{\vel}{\bm{u}}
\newcommand{\xh}{{x}_\mathrm{h}}

\newcommand{\Pd}{P_\mathrm{2D}}

\newcommand{\tcool}{t_\mathrm{cool}}

\newcommand{\bsurf}{\beta_\mathrm{surf}}
\newcommand{\bmid}{\beta_\mathrm{mid}}

\newcommand{\Qsurf}{Q_\mathrm{surf}}
\newcommand{\Qirr}{Q_\mathrm{irr}}
\newcommand{\Qrad}{Q_\mathrm{rad}}
\newcommand{\Qmid}{Q_\mathrm{mid}}
\newcommand{\Qrelax}{Q_\mathrm{relax}}
\newcommand{\Erad}{E_\mathrm{rad}}
\newcommand{\aR}{a_\mathrm{R}}
\newcommand{\lrad}{l_\mathrm{rad}}

\newcommand{\Rh}{R_\mathrm{H}}
\newcommand{\xshock}{x_\mathrm{sh}}
\newcommand{\Fdep}{F_\mathrm{dep}}

\newcommand{\pluto}{\texttt{PLUTO}}

\begin{document}

\title{How two-dimensional are planet--disc interactions? II. Radiation hydrodynamics and suitable cooling prescriptions}
\titlerunning{How 2D are planet--disc interactions? II}
\author{
	Alexandros~Ziampras\thanks{E-mail: a.ziampras@lmu.de}\inst{\ref{LMU},\ref{QMUL},\ref{MPIA}}
	\and Amelia~J.~Cordwell\inst{\ref{DAMTP}}
	\and Roman~R.~Rafikov\inst{\ref{DAMTP},\ref{IAS}}
	\and Richard~P.~Nelson\inst{\ref{QMUL}}
}

\institute{
	Ludwig-Maximilians-Universit{\"a}t M{\"u}nchen, Universit{\"a}ts-Sternwarte, Scheinerstr.~1, 81679 M{\"u}nchen, Germany\label{LMU}
    \and Astronomy Unit, Dept. of Physics and Astronomy, Queen Mary University of London, London E1 4NS, UK\label{QMUL}
	\and Max Planck Institute for Astronomy, K{\"o}nigstuhl 17, 69117 Heidelberg, Germany\label{MPIA}
	\and DAMTP, University of Cambridge, Wilberforce Road, Cambridge CB3 0WA, UK\label{DAMTP}
	\and Institute for Advanced Study, Einstein Drive, Princeton, NJ 08540, USA\label{IAS}
}

\date{\today}

\abstract
	{~The ring and gap structures found in observed protoplanetary disks are often attributed to embedded gap-opening planets and typically modeled with simplified thermodynamics and in the 2D, thin disk approximation. At the same time, both analytical and numerical studies have shown that radiative cooling and meridional processes play key roles in planet--disk interaction, though their computational cost and modeling complexity have limited their exploration.}
	{~We investigate the differences between 2D and 3D models of gap-opening planets while also comparing different thermodynamical frameworks ranging from locally isothermal to radiative hydrodynamics. We also compare simplified cooling recipes to fully radiative models in an effort to motivate the inclusion of radiative effects in future modeling even in a parametrized manner.}
	{~We perform high-resolution hydrodynamical simulations in both 2D and 3D, and then compare the angular momentum deposition by planetary spirals between the two sets of models to assess gap opening efficiency. We repeat comparisons with different thermodynamical treatments: locally isothermal, adiabatic, local $\beta$ cooling, and fully radiative including radiative diffusion.}
	{~We find that 2D models are able to capture the essential physics of gap opening with remarkable accuracy, even when including full radiation transport in both cases. Simple cooling prescriptions can capture the trends found in fully radiative models, albeit slightly overestimating gap opening efficiency near the planet. Inherently 3D effects such as vertical flows that cannot be captured in 2D can explain the differences between the two approaches, but do not impact gap opening significantly.}
	{~Our findings encourage the use of models that include radiative processes in the study of planet--disk interaction, even with the simplified yet physically motivated cooling prescriptions that we provide in lieu of full radiation transport. This is particularly important in the context of substructure-inducing planets in the disk regions where instruments such as ALMA are sensitive ($\gtrsim10$\,au).}


\bibpunct{(}{)}{;}{a}{}{,}

\maketitle

\section{Introduction}
\label{sec:introduction}

Several recent observational efforts have demonstrated that substructure in the form of rings and gaps in continuum emission is ubiquitous in disks around $\sim$Myr-old young stellar objects \citep[e.g.,][]{alma-etal-2015a,andrews-etal-2018,teague-etal-2025}. While several different mechanisms have been proposed to explain such substructure, young exoplanets present a particularly attractive scenario as the planet formation process likely begins quite early in these systems \citep{birnstiel-2024,kleine-etal-2020} and many of the observed features are well-reproduced with models of planet--disk interaction \citep[e.g.,][]{zhang-etal-2018,bae-etal-2019,toci-etal-2020}.

With the number of confirmed observed planets during the disk phase standing at three \citep{keppler-etal-2018,haffert-etal-2019,vanCapelleveen2025} theoretical modeling is by and large the main avenue to studying planet--disk interaction. A young planet interacts gravitationally with its surrounding disk, driving spiral density waves that propagate through the disk  \citep{goldreich-tremaine-1979,rafikov-2002a,ogilvie-lubow-2002}. As they travel, these density waves evolve due to nonlinear effects, eventually turning into shocks, delivering angular momentum to the disk and carving gaps around the planet's orbit \citep{goodman-rafikov-2001,rafikov-2002b}.

Models of gap-opening planets have increased in sophistication over the years, highlighting the effects of radiative cooling \citep{miranda-rafikov-2020a,miranda-rafikov-2020b, ziampras-etal-2020b,ziampras-etal-2023a}, dust dynamics \citep{weber-etal-2019,bi-etal-2021,bi-etal-2023}, and even the dependence of cooling on the local dust content \citep{binkert-etal-2023,krapp-etal-2024,ziampras-etal-2025}. In particular, it has been shown that planet-driven gap opening is highly sensitive to the cooling timescale \citep{miranda-rafikov-2020a,zhang-zhu-2020,zhang-etal-2024} as well as radiative diffusion across the planetary spiral shocks \citep{miranda-rafikov-2020b,ziampras-etal-2023a}. Combined with the complicated opacity models in literature \citep{semenov-etal-2003,birnstiel-etal-2018,woitke-etal-2016}, surveying a wide parameter space while maintaining accuracy when attempting to model an observed system with a substructure-forming planet can be a lost cause.

While the aforementioned models are of key importance to understanding planet--disk interaction, they are also both sensitive to several---often poorly constrained---physical parameters (e.g., gas density, dust composition, dust-to-gas ratio) and computationally expensive, such that sweeping the parameter space with 3D models is unfeasible. To that end, vertically integrated, 2D models are quite attractive, as the lower computational cost enables the execution of large parameter studies at high resolution.

Of course, the question remains whether such 2D models can accurately capture the physics of gap opening in the first place. The flow around a gap-opening planet is inherently three dimensional, featuring zonal flows and meridional circulation \citep{Morbidelli2014,fung-chiang-2016,bi-etal-2021,wafflard-fernandez-lesur-2023}. At the same time, while extensive work has being carried out to verify that planet--disk interaction can be mapped from 3D to 2D \citep[e.g.,][]{mueller-etal-2012,fung-chiang-2016,lega-etal-2021,hammer-lin-2023,zier-springel-2023,brown-ogilvie-2024,cordwell-etal-2025}, radiative processes complicate this picture as they are very sensitive to the optical depth and cooling timescale, both of which are strong functions of $z$ \citep[e.g.,][]{bae-etal-2021,ziampras-etal-2024b}. As such, a comparison of 2D to 3D models while including radiation hydrodynamics is necessary to determine whether and under what conditions the 2D, vertically integrated cooling prescriptions can be reconciled with fully 3D models.

In this work, we perform 2D and 3D radiation hydrodynamical simulations of planet--disk interaction comparing the gap-opening efficiency of the two frameworks for the same physical parameters. We re-use some of the 2D models in \citet{ziampras-etal-2023a} but re-analyze them with a more appropriate method following \citet{cordwell-rafikov-2024}, and supplement them with fully 3D simulations which we analyze following \citet{cordwell-etal-2025}. Our goals are to both determine whether 2D models can accurately capture the gap-opening efficiency of 3D models, and to identify cooling prescriptions that can approximate the behavior of fully radiative models with good accuracy.

The paper structure is as follows: in Sect.~\ref{sec:physics-numerics} we detail our physical and numerical frameworks in both 2D and 3D, and define effective cooling timescales that we found to be useful in approximating the behavior of fully radiative models. We also define our simulation metrics which we use to compare the two frameworks. In Sect.~\ref{sec:results} we present our results, where we compare the gap-opening efficiency of 2D and 3D models. We then discuss the implications of our findings in Sect.~\ref{sec:discussion}, and conclude in Sect.~\ref{sec:summary}.

\section{Physics and numerics}
\label{sec:physics-numerics}

In this section we describe our physical framework and numerical approach. While both are largely similar to \citet{ziampras-etal-2023a} and \citet{ziampras-etal-2024b} for 2D and 3D respectively, we reiterate their description here for clarity and to facilitate the derivation of equivalent cooling timescales between 2D and 3D in Sect.~\ref{sub:theory-beta}.

\subsection{Three dimensional framework}
\label{sub:theory-3D}

We consider an inviscid disk of gas with volume density $\rho$, internal energy density $e$ and velocity field $\vel$ around a star with mass $\Mstar=1\,\Msun$ and luminosity $\Lstar=1\,\Lsun$. The gas has a mean molecular weight $\mu=2.353$ and an adiabatic index $\gamma=7/5$. The inviscid Navier--Stokes equations then read \citep[e.g.,][]{tassoul-1978}
\begin{subequations}
	\label{eq:navier-stokes}
	\begin{align}
		\label{eq:navier-stokes-1}
		\DP{\rho}{t} + \vel\cdot\nabla\rho=-\rho\nabla\cdot\vel,
	\end{align}
	\begin{align}
	\label{eq:navier-stokes-2}
		\DP{\vel}{t}+ (\vel\cdot\nabla)\vel=-\frac{1}{\rho}\nabla P -\nabla\Phi,
	\end{align}
	\begin{align}
	\label{eq:navier-stokes-3}
		\DP{e}{t} + \vel\cdot\nabla e=-\gamma e\nabla\cdot\vel + Q.
	\end{align}
\end{subequations}
Here, the pressure is given by the perfect gas equation of state with $P = (\gamma-1)e$. Through the latter we can define the isothermal sound speed $\cs=\sqrt{P/\rho}$, the temperature $T=\mu\cs^2/\Rgas$, the pressure scale height $H=\cs/\OmegaK$, and the aspect ratio $h = H/R$, with $\OmegaK = \sqrt{\G\Mstar/R^3}$ being the Keplerian angular speed at cylindrical distance $R$ from the star. The perfect gas and gravitational constants are $\Rgas$ and $\G$, respectively. Finally, $\Phi=\Phi_\star + \Phi_\text{p}$ is the gravitational potential of the star and planet with mass $\Mp$ and radial location $\Rp$, given by $\Phi_\star = -\G\Mstar/r$ and
\begin{equation}
\label{eq:planet-potential}
\Phi_\mathrm{p} = \begin{cases}
	-\G\Mp/d, & d \geq \epsilon \\
	-\frac{\G\Mp}{\epsilon}\left[\frac{d^3}{\epsilon^3} - 2\frac{d^2}{\epsilon^2}+ 2\right], & d < \epsilon,
\end{cases}
\end{equation}
in order to avoid singularities, following \citet{klahr-kley-2006}. Here, $\epsilon=0.5\,\Rh = 0.5\,\Rp\sqrt[3]{\Mp/3\Mstar}$, and $\bm{d}=\bm{r}-\bm{R}_\mathrm{p}$ is the distance between a point at position $\bm{r}$ and the planet. This choice of smoothing length $\epsilon$, while larger than that used in \citet{cordwell-etal-2025}, is still small enough to not affect the flow in our region of interest. We account for the indirect acceleration arising from the star--planet system orbiting their mutual center of mass.

In Eq.~\eqref{eq:navier-stokes-3}, $Q$ represents additional source terms that can affect the thermal energy evolution. We define four different thermodynamical treatments similar to \citet{ziampras-etal-2023a}:
\begin{itemize}
	\item \emph{locally isothermal}: Eq.~\eqref{eq:navier-stokes-3} is ignored altogether, and a temperature profile $T_0(\bm{r})$ is prescribed such that $P=\Rgas\rho T_0/\mu$. This is the most simplistic approach within our framework, effectively translating to immediate thermal relaxation;
	\item[]
	\item \emph{adiabatic}: Eq.~\eqref{eq:navier-stokes-3} is evolved with $Q=0$. This approach captures adiabatic compression while conserving entropy (except at shock fronts), but assumes no thermal relaxation;
	\item[]
	\item \emph{local cooling}: thermal relaxation is parametrized with a cooling timescale $\tcool=\beta\OmegaK^{-1}$, such that in Eq.~\eqref{eq:navier-stokes-3}
	\begin{equation}
		\label{eq:Qrelax}
		Q = \Qrelax = -f_\beta\frac{e-e_0}{\tcool},\quad e_0=\frac{\Rgas}{\mu(\gamma-1)}\rho T_0 = \cv\rho T_0,
	\end{equation}
	with $f_\beta$ being a prefactor that differs from unity for temperature-dependent $\beta$ \citep{dullemond-etal-2022,ziampras-etal-2023a};
	\item[]
	\item \emph{fully radiative}: the disk cools by interacting with the local radiation field $\Erad$ in the flux-limited diffusion approximation \citep[FLD,][]{levermore-pomraning-1981}. Here, $\Erad$ is dynamically evolved and added to the set of equations in Eq.~\eqref{eq:navier-stokes} with
	\begin{equation}
		\label{eq:Erad}
		\DP{\Erad}{t} + \nabla\cdot\bm{F}=-\Qrad,\quad \bm{F} = -\frac{\lambda c}{\kappaR\rho}\nabla\Erad,
	\end{equation}
	and the matter--radiation field coupling is given by defining in Eq.~\eqref{eq:navier-stokes-3}
	\begin{equation}
		\label{eq:Qrad}
		Q = \Qrad = -\kappaP\rho c\left(\aR T^4 - \Erad\right).
	\end{equation}
	In this context, $\kappaR$ and $\kappaP$ are the Rosseland and Planck mean opacities, $c$ is the speed of light, $\aR$ is the radiation constant, and $\lambda$ is a flux limiter following \citet{kley-1989} that captures the transition between the optically thick diffusion limit ($\lambda\rightarrow1/3$) and the optically thin, free-streaming limit ($F\rightarrow c\Erad$).	This is the most realistic approach within our framework, capturing both the thermal relaxation of $e$ through $\Qrad$ and the radiative diffusion of $\Erad$ through Eq.~\eqref{eq:Erad}. To prevent the disk from cooling indefinitely, we prescribe $\Erad=\aR T_0(R)^4$ at the surfaces and radial ends of the disk as in \citet{ziampras-etal-2024b}, assuming that the disk is heated by reprocessed starlight from the disk surfaces.
\end{itemize}

\subsection{Two dimensional framework}
\label{sub:theory-2D}

Similar to the set of equations in Eq.~\eqref{eq:navier-stokes} and by defining the surface density $\Sigma = \int_{-\infty}^{\infty}\rho\mathrm{d}z$ and vertically integrated pressure $\Pd = \Sigma\cs^2$, (assuming $T$ is independent of $z$) the inviscid Navier--Stokes equations in the thin disk approximation read
\begin{subequations}
	\label{eq:navier-stokes-2D}
	\begin{align}
		\label{eq:navier-stokes-2D-1}
		\DP{\Sigma}{t} + \vel\cdot\nabla\Sigma=-\Sigma\nabla\cdot\vel,
	\end{align}
	\begin{align}
	\label{eq:navier-stokes-2D-2}
		\DP{\vel}{t}+ (\vel\cdot\nabla)\vel=-\frac{1}{\Sigma}\nabla \Pd -\nabla\Phi,
	\end{align}
	\begin{align}
	\label{eq:navier-stokes-2D-3}
		\DP{e}{t} + \vel\cdot\nabla e=-\gamma e\nabla\cdot\vel + Q.
	\end{align}
\end{subequations}
In the above equations the thermal energy density has been redefined as $e=\Pd/(\gamma-1)$. To capture the vertical structure of the disk, the gravitational potential of the planet is also replaced by a Plummer potential with
\begin{equation}
	\Phi_\mathrm{p} = -\frac{\G\Mp}{\sqrt{d^2+\epsilon^2}}, \quad \epsilon = 0.6\,\Hp,
\end{equation}
following \citet{mueller-etal-2012}.
While \citet{cordwell-etal-2025} showed that the newer potential form of \cite{brown-ogilvie-2024} functions more optimally than the traditional Plummer potential, the main purpose of this paper is to explore the role of disk thermodynamics rather than the planetary potential on gap opening. As we seek to identify how well the overall morphology, rather than quantitative details, of gap opening can be represented in 2D simulations, the Plummer approximation is an appropriate choice.

While $\Qrelax$ maintains its form as in Eq.~\eqref{eq:Qrelax} within the local cooling approach (with $\Sigma$ instead of $\rho$), the radiative cooling term $\Qrad$ in Eq.~\eqref{eq:Erad} is now split among a vertical, surface cooling component $\Qsurf$, an in-plane radiative diffusion term $\Qmid$, and an irradiation heating term $\Qirr$. The two cooling terms can be written as
\begin{equation}
	\label{eq:Qsurf}
	\Qsurf = -2\frac{\sigmaSB T^4}{\taueff},~~~~\taueff = \frac{3\tauR}{8} + \frac{\sqrt{3}}{4} + \frac{1}{4\tauP},~~~~\tau_{\mathrm{R,P}} = \frac{1}{2}\kappa_\mathrm{R,P}\Sigma
\end{equation}
for surface cooling following the effective optical depth model of \citet{hubeny-1990}, and
\begin{equation}
	\label{eq:Qmid}
	\Qmid = \sqrt{2\pi}H\,\nabla\cdot\left(\frac{16\lambda \sigmaSB T^3}{\kappaR\rhomid}\nabla T\right),\quad\rhomid\equiv \frac{1}{\sqrt{2\pi}}\frac{\Sigma}{H}
\end{equation}
for in-plane radiative diffusion, where we further assume that $\Erad=\aR T^4$ at the disk midplane (one-temperature approximation). In the above, $\sigmaSB$ is the Stefan--Boltzmann constant and $\rhomid$ is the volume density at the disk midplane, obtained assuming vertical hydrostatic equilibrium.

Finally, we include the prescription of irradiation heating by \citet{menou-goodman-2004} to balance surface losses via Eq.~\eqref{eq:Qsurf}:
\begin{equation}
	\label{eq:Qirr}
	\Qirr = 2\frac{\Lstar}{4\pi R^2}(1-\varepsilon)\left(\D{\log H}{\log R}-1\right)\frac{h}{\taueff},
\end{equation}
with $\D{\log H}{\log R} = 9/7$ and a disk albedo $\varepsilon=1/2$.

\subsection{Effective cooling timescales}
\label{sub:theory-beta}

One of the goals of this work is to identify cooling timescale prescriptions that can approximate the behavior of fully radiative models with good accuracy, capturing radiative effects through local thermal relaxation (Eq.~\eqref{eq:Qrelax}) rather than solving for the full radiative diffusion problem (Eqs.~\eqref{eq:Erad}~\&~\eqref{eq:Qrad}). Following previous work \citep{flock-etal-2017a,ziampras-etal-2024b} we define the cooling timescale within the 3D framework as
\begin{equation}
	\label{eq:beta-3D}
	\beta^\mathrm{3D} = \frac{\OmegaK}{\eta}\left(H^2 + \frac{\kappaP}{\kappaR}\frac{\lrad^2}{3}\right),\quad\eta = \frac{16\sigmaSB T^3}{3\kappaR\rho^2\cv}, \quad \lrad=\frac{1}{\kappaP\rho},
\end{equation}
where $\eta$ and $\lrad$ are the radiative diffusion coefficient and photon mean free path, respectively. This recipe captures cooling in both the optically thick, diffusion-limited regime (with the assumption that the diffusion length scale is $\sim H$) and in the optically thin regime where the cooling rate is proportional to the emissivity $\kappaP$. We note that the factor $\kappaP/\kappaR$ is a correction to the formulas provided by the aforementioned studies that arises when relaxing the assumption that $\kappaR=\kappaP=\kappa$. A heuristic derivation of Eq.~\eqref{eq:beta-3D} is provided in Appendix~\ref{apdx-beta}.

In the 2D framework, the above formula is still relevant as it represents cooling via diffusion through the disk plane (defining all parameters at the midplane), effectively capturing the effects of $\Qmid$ in Eq.~\eqref{eq:Qmid}. To further include the effects of surface cooling, assuming that heat is removed locally along the $z$ direction, we can write the two components of surface and in-plane cooling as
\begin{equation}
	\label{eq:bmid-bsurf-2D}
	\bmid = \beta^\text{3D}(z=0),\qquad\bsurf = \frac{e}{|\Qsurf|}\OmegaK.
\end{equation}
The two can then be combined such that cooling is always limited by the fastest of the two channels \citep[e.g.,][]{miranda-rafikov-2020b} to yield an effective cooling timescale:
\begin{equation}
	\label{eq:beta-2D}
	\beta^\mathrm{2D} = \left[\bmid^{-1} + \bsurf^{-1}\right]^{-1}.
\end{equation}
Similar to \citet{ziampras-etal-2023a}, Eq.~\eqref{eq:beta-2D} can be algebraically rearranged to express $\beta^\text{2D}$ as
\begin{equation}
	\label{eq:beta-2D-f}
	\beta^\mathrm{2D} = \frac{1}{f+1}\bsurf = \frac{1}{f+1}\frac{e}{|\Qsurf|}\OmegaK,\quad
	f = \frac{16\pi\,\tauP\taueff}{6\tauR\tauP + \pi},  
\end{equation}
with $f\rightarrow\pi$ and $f\rightarrow 4$ in the optically thick and thin limits ($\tau\rightarrow\infty$ and $\tau\rightarrow0$, respectively) if $\tauR=\tauP=\tau$.

We note that Eqs.~\eqref{eq:beta-3D}~\&~\eqref{eq:beta-2D} can be used to both estimate the cooling timescale in radiative models and compute a relaxation rate $\Qrelax$ when using a local cooling approach. In the latter case, \citet{dullemond-etal-2022} showed that the prefactor in Eq.~\eqref{eq:Qrelax} takes the form $f_\beta=1/(4\pm b)$ in the optically thick \mbox{(+)} and thin \mbox{(-)} limits, where $\taueff\propto T^{\pm b}$ for temperature-dependent opacities (see Eq.~\eqref{eq:Qsurf}). Nevertheless, in the range $\tau\sim0.1$--10, where $\taueff$ is rather flat due to the constant term $\sqrt{3}/4$ in Eq.~\eqref{eq:Qsurf}, this simplifies to $f_\beta\approx1/4$ \citep{ziampras-etal-2023a}. We use this value of $f_\beta$ for our models with local cooling. In Sect.~\ref{sec:results}, we will show that models using this local cooling approach with our analytic formulas in Eqs.~\eqref{eq:beta-3D}~\&~\eqref{eq:beta-2D} perform very well when compared to their fully radiative counterparts in the context of planet-driven gap opening.

\subsection{Numerical setups}
\label{sub:numerics}

We use the numerical (magneto)hydrodynamics finite-volume code \pluto{} \citep{mignone-etal-2007} in 3D spherical $\{r,\theta,\phi\}$ or 2D cylindrical $\{R,\phi\}$ polar geometry, depending on framework. We use the flux-limited diffusion radiation transport modules detailed in \citet{ziampras-etal-2024b} and \citet{ziampras-etal-2020a} for our 3D and 2D runs, respectively\footnote{For FLD, we use the \texttt{PetSc} library \citep{petsc-web-page} in 3D, and a simple successive over-relaxation technique in 2D.}. The models are run in a frame corotating with the planet's orbital speed $\Omega_\text{p}=\sqrt{\G(\Mstar+\Mp)/\Rp^3}$ and make use of the FARGO transport algorithm \citep{masset-2000} implemented in \pluto{} by \citet{mignone-etal-2012} to both minimize numerical diffusion and alleviate the strict timestep limitation of the rapidly rotating inner radial boundary. Finally, we employ a third-order weighted essentially non-oscillatory reconstruction scheme \citep[WENO,][]{yamaleev-carpenter-2009} and third-order Runge--Kutta timestepping, the slope limiter by \citet{vanleer-1974}, and the HLLC Riemann solver \citep{toro-etal-1994}.

Our numerical grid spans $R$ or $r\in[0.4,2.5]\,\Rp$ with logarithmic radial spacing, covers the full $\phi\in[0,2\pi]$ in azimuth, and, in our 3D models, extends between $\theta\in[\pi/2-0.15,\pi/2]$ in the polar direction. For our choice of temperature profile with $h=0.05\,(R/\Rp)^{1/4}$ (see text below), this translates to $z_\mathrm{max}=3H$ at $R=\Rp$.
Our boundaries are periodic in $\phi$, symmetric about the midplane, and we impose wave-damping regions for $R<0.5\,\Rp$, $R>2.1\,\Rp$ following \citet{devalborro-etal-2006} with a damping timescale of 0.1 local orbits.

After carrying out a resolution study (see Appendix~\ref{apdx:resolution}), we choose a grid resolution of $N_R\times N_\phi = 1200\times4096$ and $N_r\times N_\theta\times N_\phi = 600\times48\times2048$ cells for our 2D and 3D models, which translates to 32 and 16 cells per scale height, respectively. This choice does not sacrifice quality for performance, however, but rather shows that our 3D models converged within 95\% for half the effective resolution compared to our 2D runs with respect to our metrics of choice.

Similar to \citet{ziampras-etal-2023a}, we aim to build a set of models where $\tau$, $\Sigma$, and $\beta$ are constant within our simulation domain (for 3D models, we seek to satisfy this at the midplane). This is achieved by setting $\kappaR=\kappaP=\kappa=\text{const.}$, $\Sigma_0=\Sigma(t=0)=\text{const.}$, and $T_0 = T(t=0) = T_\text{p}\,(R/\Rp)^{-1/2}$. For a solar-type star, thermal equilibrium between $\Qsurf$ and $\Qirr$ in Eqs.~\eqref{eq:Qsurf}~\&~\eqref{eq:Qirr} translates to $T_\text{p}=21$\,K at $\Rp=30$\,au, or $\hp=0.05$. We then use an iterative approach to compute $\kappa$ and $\Sigma_0$ to achieve different values of $\beta^\text{2D}$, or $\beta^\text{3D}$ at the disk midplane while maintaining an optical depth $\tau=\frac{1}{2}\kappa\Sigma=10$. In sections where we compare our findings as a function of the cooling timescale, we use $\beta^\text{2D}$ as a reference value for the disk cooling time, regardless of the equation of state. This is done because it represents the total cooling timescale in our 2D runs, whereas in 3D the effective cooling timescale might be much shorter than $\beta^\text{3D}(z=0)$ due to the entire column from the midplane to the vertical boundary contributing to cooling.

We note that the actual values of $\kappa$ and $\Sigma_0$, however unrealistic, are completely arbitrary. The goal is to establish numerical experiments where $\tau$ and $\beta$ are (initially) constant throughout the disk midplane in order to measure the dissipation of planet-driven spirals as a function of the cooling timescale while maintaining a marginally optically thick disk ($\tau_\text{mid}=10$). While this is trivial for models with local cooling, where the value of $\beta$ can be simply prescribed, running equivalent radiative models requires defining our initial conditions in this way such that $\beta$ evaluates to its target value through Eq.~\eqref{eq:beta-3D}~or~\eqref{eq:beta-2D} in 3D and 2D, respectively.

Finally, we choose a planet mass of $\Mp=3.75\times 10^{-5}\Mstar = 12.5\,\Mearth$. For our choice of parameters, we define the thermal mass following \citep{goodman-rafikov-2001} as $\Mth = \frac{2}{3} h^3 \Mstar \approx 28\,\Mearth$, such that $\Mp = 0.42\,\Mth$\footnote{We note that this definition of the thermal mass is slightly different from that of $\Mth=h^3\Mstar$ in e.g. \citet{cordwell-etal-2025}.}.
This places the planet in a quasi-nonlinear regime, where it is expected to carve a gap due to the lack of viscous diffusion but its spiral wakes will not steepen into shocks before traveling a radial distance of \citep{goodman-rafikov-2001}
\begin{equation}
	\label{eq:xshock}
	\xshock \approx 0.93\Hp \left(\frac{\gamma+1}{12/5} \frac{\Mp}{\Mth}\right)^{-2/5}
\end{equation}
away from the planet. This allows us to measure differences in gap opening between isothermal, adiabatic, and radiative models, as no angular momentum should be deposited in the region $|R-\Rp|<\xshock$ in the absence of radiative damping \citep{rafikov-2002a}. The planet is ramped to its full mass during its first orbit in the simulation following the formula in \citet{devalborro-etal-2006}. This ensures that the spiral patterns have been established throughout the computational domain and have reached a steady state within a few ($\sim\!5$) planetary orbits.

\subsection{Simulation metrics}
\label{sub:metrics}

As previous studies have already shown, the gap opening process is a sensitive function of the cooling timescale. To quantify this effect, we will focus on the planet's primary gap, even though multiple gaps can in principle be opened in inviscid disks \citep[e.g.,][]{zhang-etal-2018}. We therefore need a metric to measure the efficiency with which the planet removes material from its local orbit. Rather than computing an ``angular momentum budget'' as in \citet{ziampras-etal-2023a}, who integrated the total spiral angular momentum flux $F_J$ throughout the disk, we will instead compute the angular momentum deposition function $\partial \Fdep/\partial R$ following \citet{cordwell-rafikov-2024}. This can be defined as
\begin{equation}
	\label{eq:fdep}
	\DP{\Fdep}{R} = \DP{T}{R} - \DP{F_J}{R},
\end{equation}
where the angular momentum flux $F_J$ and gravitational torque density $\partial T/\partial R$ are given in the 2D framework by
\begin{subequations}
	\label{eq:FJ-T-2D}
	\begin{align}
		\label{eq:FJ-2D}
		F_J(R) = R^2 \oint \Sigma \delta u_R \delta u_\phi \mathrm{d}\phi,\quad\delta u_x = u_x - \langle u_x\rangle,
	\end{align}
	\begin{align}
		\label{eq:T-2D}
		\DP{T}{R} = R \oint \Sigma \DP{\Phi_\mathrm{p}}{\phi} \mathrm{d}\phi,
	\end{align}
\end{subequations}
and in the 3D framework computed after first integrating along $z$ \citep{cordwell-etal-2025}:
\begin{subequations}
	\label{eq:FJ-T-3D}
	\begin{align}
		\label{eq:FJ-3D}
		F_J(R) = R^2 \oint \left[\int_{-\infty}^{\infty} \rho \delta u_R \delta u_\phi \mathrm{d}z \right] \mathrm{d}\phi,
	\end{align}
	\begin{align}
		\label{eq:T-3D}
		\DP{T}{R} = R \oint \left[\int_{-\infty}^{\infty} \rho \DP{\Phi_\mathrm{p}}{\phi} \mathrm{d}z \right] \mathrm{d}\phi.
	\end{align}
\end{subequations}
In the above equations, $\langle u_x\rangle$ denotes density-weighted azimuthal averages such that
\begin{equation}
	\label{eq:sigma-avg}
	\langle u_x \rangle = \frac{1}{2\pi\bar{\rho}}\oint u_x \rho \mathrm{d}\phi
\end{equation}
with $\bar{\rho}$ being the azimuthally averaged density ($\rho$ and $\bar{\rho}$ are replaced with $\Sigma$ and $\bar{\Sigma}$ in 2D). After computing $\partial \Fdep/\partial R$, we can then reconstruct the azimuthally averaged perturbed density $\Delta\bar{\Sigma}/\Sigma_0$ as a function of time by computing the surface density evolution rate $\partial\Sigma/\partial t$ using $\partial \Fdep/\partial R$, following Eqs.~(C1)--(C12) in Appendix~C of \citet{cordwell-rafikov-2024}\footnote{Tools for the calculation of $\Delta\Sigma/\Sigma_0$ from $\partial \Fdep/\partial R$ are available at \href{https://github.com/cordwella/disc\_planet\_analysis}{https://github.com/cordwella/disc\_planet\_analysis}.}:
\begin{equation}
	\label{eq:sdot}
	\frac{\Delta\bar{\Sigma}}{\Sigma_0}(t)\approx t\,\frac{1}{\Sigma_0}\frac{\partial\Sigma}{\partial t}.
\end{equation}
Here, $\Sigma_0$ is the initial surface density profile. We note that this approach assumes that the gap opening process operates in its linear regime, where $\partial\Sigma/\partial t$ is constant in time. This is a very accurate approximation for gap depths less than $20\%$ \citep{cordwell-rafikov-2024}, a condition met by the circumstances considered in this paper. As we are interested in the inviscid regime of gap opening we are unable to make direct comparisons to standard viscous steady state models such as in \citet{kanagawa-etal-2015b}, \citet{zhang-etal-2018}, or \citet{duffell-2020}.

In all figures shown below, $\partial \Fdep/\partial R$ is normalized to
\begin{align}
    \label{eq:FJ0}
    \frac{F_{J,0}}{\Rp} = \left(\frac{\Mp}{M_\star}\right)^2 \hp^{-3} \Sigma_\mathrm{p} \Rp^3 \Omega_\mathrm{p}^2,
\end{align}
where ${F_J}_0$ is the characteristic one-sided torque due to the planet \citep{goldreich-tremaine-1980}. 

For our analysis in the following sections, we will focus on the angular momentum deposition due to spiral shocks rather than dynamics within the planet's corotating region. The latter should not contribute to spiral-driven gap opening in the absence of radiative damping. However, dynamical processes such as the vertical shear instability \citep[VSI,][]{nelson-etal-2013,stoll-etal-2017b,ziampras-etal-2023vsi}, buoyancy modes \citep{zhu-etal-2012,mcnally-etal-2020}, or transient features while the horseshoe region is undergoing phase mixing (e.g., the vortensity striping seen in \citet{cordwell-etal-2025}) can still drive correlated velocity fluctuations in that region.
In particular, the VSI will operate for very short cooling timescales \citep[$\beta\lesssim0.1$,][]{lin-youdin-2015}, while the disk's buoyancy response can drive substantial motion for nearly adiabatic conditions \citep[$\beta\gg100$,][]{yun-etal-2022,ziampras-etal-2024b}. To avoid the effects of either of these processes, we set $\partial \Fdep/\partial R=0$ within the planet's corotating region $|R-\Rp|<\xh$, with the horseshoe half-width $\xh$ defined following \citet{paardekooper-etal-2010}:
\begin{equation}
	\label{eq:xhorse}
	\xh = \frac{1.1}{\gamma^{1/4}}\left(\frac{0.4}{\epsilon/H}\right)^{1/4} \sqrt{\frac{\Mp}{h\Mstar}}\Rp,\quad \epsilon=0.6\,\Hp.
\end{equation}
The radial profile of $\partial \Fdep/\partial R$ is then smoothed with a rolling average over a width of $H$ to eliminate any numerical noise, and $\partial \Sigma/\partial t$ is computed using this smoothed data. We found that this smoothing has a negligible effect on the results, but improves the readability of the figures. Nevertheless, we show the raw data in faded colors in Figs.~\ref{fig:gap-view}--\ref{fig:sdot-eos} for transparency.

\section{Results}
\label{sec:results}

In this section we present the results of our numerical experiments. We begin with a demonstration of our approach, continue to an analysis of the angular momentum deposition function and the resulting gap structures as a function of the cooling timescale, and conclude with a comparison between our 2D and 3D models.

\subsection{A fiducial model}
\label{sub:results-fiducial}

\begin{figure*}
	\centering
	\includegraphics[width=\textwidth]{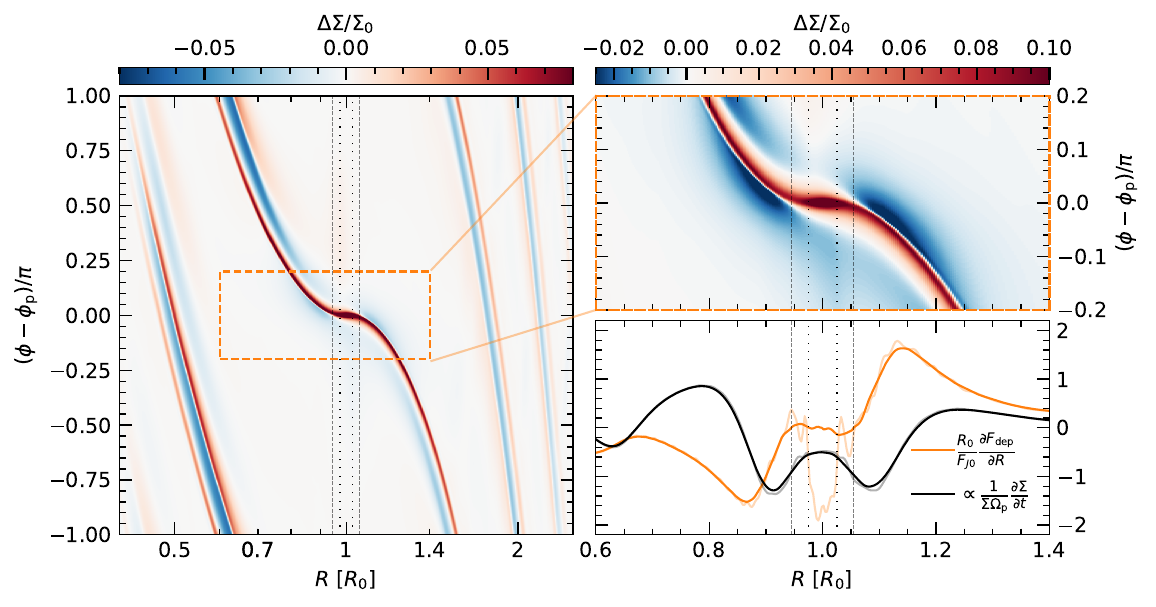}
	\caption{Snapshot of the perturbed surface density after 10 planetary orbits for our locally isothermal 3D model. Left: the entire simulation domain. Spiral shocks are weaker near either radial end of the disk due to wave damping. Top right: a zoom-in into the orange box on the left, highlighting the region where gap opening is driven. Bottom right: the angular momentum deposition function $\partial \Fdep/\partial R$ and the surface density evolution rate $\partial\Sigma/\partial t$ in arbitrary units computed according to \citet{cordwell-rafikov-2024}. Saturated and faded colors indicate smoothed or raw data, respectively. Dashed and dotted vertical lines mark the isothermal shock distance $\xshock$ (Eq.~\eqref{eq:xshock} with $\gamma=1$) and the planet's corotating region (Eq.~\eqref{eq:xhorse} with $\gamma=1$), respectively.}
	\label{fig:gap-view}
\end{figure*}

To illustrate our approach, we show a snapshot of the surface density structure after 10 planetary orbits for our locally isothermal 3D model in Fig.~\ref{fig:gap-view}. On the left panel we show the entire simulation domain, with the planet located at $R=R_0$ and spirals permeating the disk. The top right panel shows a zoomed-in view of the region around the planet (indicated with an orange box on the left), highlighting the density perturbations around the spiral shocks. The bottom right panel shows the angular momentum deposition function $\partial \Fdep/\partial R$ in normalized units, and the surface density evolution rate in arbitrary units computed based on this $\partial \Fdep/\partial R$ according to \citet{cordwell-rafikov-2024}. Dashed vertical lines mark the shock distance $\xshock$ following Eq.~\eqref{eq:xshock}, and dotted lines mark the edge of the planet's corotating region $\xh$ following Eq.~\eqref{eq:xhorse}.

From the bottom right panel of Fig.~\ref{fig:gap-view} we can identify several key features of planet-driven gap opening by noticing that the angular momentum deposition function is---within noise levels---zero between the edge of the planet's corotating region and the shock distance. This is expected, as the spirals have not yet steepened into shocks and travel as linear waves within this region.\footnote{Horseshoe dynamics leads to a nonzero $\partial \Fdep/\partial R$ within $\xh$ (see Eq.~\eqref{eq:xhorse}), but as mentioned above we set this quantity to zero in this region.} Beyond $\xshock$, however, we see a rise in $\partial \Fdep/\partial R$ and a corresponding increase in the surface density evolution rate $\partial \Delta\Sigma/\partial t$. The resulting ``double-trough'' gap shape is entirely consistent with expectations from literature for low-viscosity, locally isothermal models \citep[e.g.,][]{rafikov-2002b,miranda-rafikov-2020a,zhang-etal-2024, cordwell-rafikov-2024}.

\begin{figure}
	\centering
	\includegraphics[width=\columnwidth]{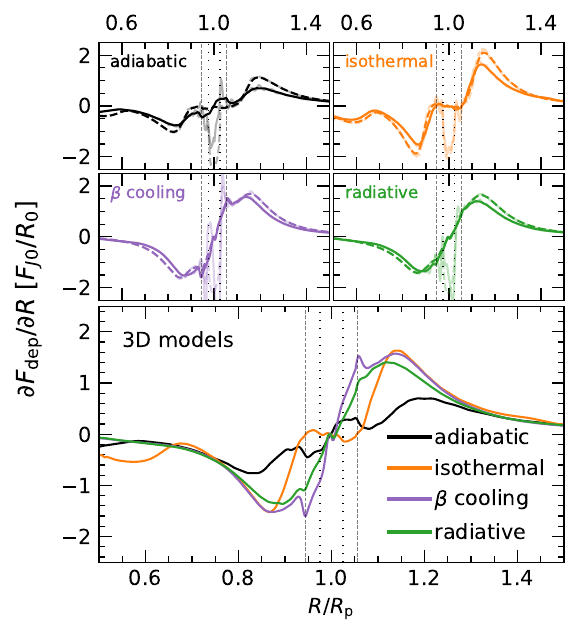}
	\caption{Angular momentum deposition $\partial \Fdep/\partial R$ for our fiducial 2D and 3D models (dashed and solid lines, respectively) and for four different equations of state (EOS). Models with cooling ($\beta$, radiative) have $\beta^\mathrm{2D}\approx2$. As in Fig.~\ref{fig:gap-view}, faded and saturated colors indicate raw and smoothed data, respectively, showing that our smoothing procedure has negligible effects beyond the corotating region. The bottom panel compares the four EOS for our 3D models, highlighting the role of cooling.}
	\label{fig:fdep-eos}
\end{figure}

\begin{figure}
	\centering
	\includegraphics[width=\columnwidth]{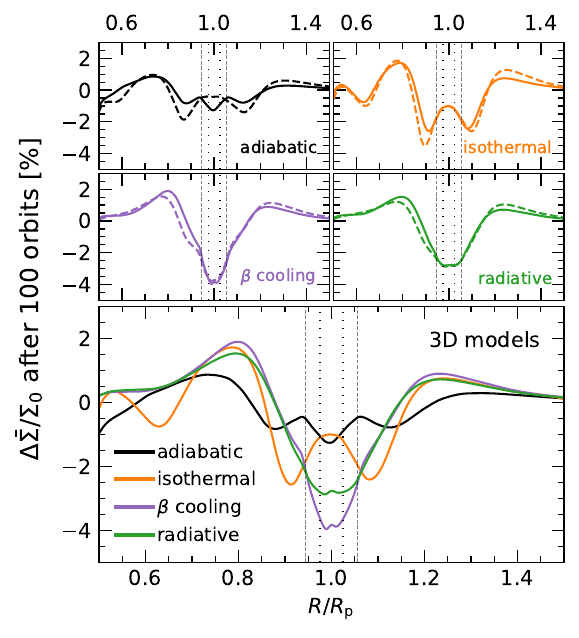}
	\caption{Estimates of the perturbed surface density expected after 100 planetary orbits for the models featured in Fig.~\ref{fig:fdep-eos} \citep[using $\partial \Fdep/\partial R$ from Fig.~\ref{fig:fdep-eos}, Eq.~\eqref{eq:sdot}, and the method in][]{cordwell-rafikov-2024}. The adiabatic and isothermal models show a double-trough gap structure, while radiative models show a single, deep gap. An apparent third dip in the adiabatic model at $R=\Rp$ is related to the perturbations driven by buoyancy modes near the edge of the planet's horseshoe region.}
	\label{fig:sdot-eos}
\end{figure}

This behavior no longer necessarily holds when a finite cooling timescale is introduced, as radiative damping can now drive angular momentum deposition well before $\xshock$ is reached. This is demonstrated in Fig.~\ref{fig:fdep-eos}, where we show the angular momentum deposition function $\partial \Fdep/\partial R$ for our fiducial 2D and 3D models (dashed and solid lines, respectively) and for four different equations of state: adiabatic (black), locally isothermal (orange), $\beta$ cooling (purple), and fully radiative (green). Our models with cooling ($\beta$, radiative) have $\beta^\mathrm{2D}\approx2$. Here, it becomes clear that a finite cooling timescale leads to a nonzero angular momentum deposition within $\xshock$. In particular, by comparing the purple and orange curves in the bottom panel of Fig.~\ref{fig:fdep-eos}, we see that $\partial \Fdep/\partial R$ not only peaks much closer to the planet, but also remains high up to $\sim\!2\xshock$ away, suggesting the formation of a single, deep gap around the planet. This is fully in line with the findings of \citet{miranda-rafikov-2020a}.

In Fig.~\ref{fig:sdot-eos} we then show the expected surface density profile after 100 planetary orbits by proxy of the surface density evolution rate through Eq.~\eqref{eq:sdot} for the same models as in Fig.~\ref{fig:fdep-eos}. In this figure, and especially in the bottom panel, we can see that the gap profile is significantly different between the different models: the locally isothermal and adiabatic models (orange and black lines, respectively) feature a shallow, double-trough gap profile, while the models with cooling (purple and green lines) show a single, deep gap, which is slightly deeper for $\beta$ cooling (purple). This is once again fully consistent with findings from previous studies \citep{miranda-rafikov-2020b,zhang-zhu-2020,zhang-etal-2024}.

We note that the profiles shown in Fig.~\ref{fig:sdot-eos} are not the result of dynamical evolution over 100 planetary orbits, but rather extrapolations using the surface density evolution rates computed using $\partial\Fdep/\partial R$ and the framework highlighted in \citet{cordwell-rafikov-2024}~\&~\citet{cordwell-etal-2025}. Nevertheless, the aforementioned studies have shown that this approximation remains valid during the early phases of gap opening (or $\Delta\Sigma/\Sigma_0$ of a few percent). We refer the reader to these studies for further details.

\subsection{Gap opening as a function of cooling timescale}
\label{sub:results-beta}

\begin{figure}
	\centering
	\includegraphics[width=\columnwidth]{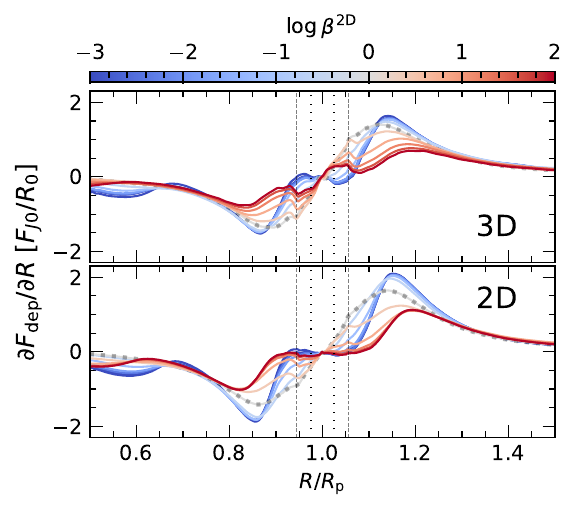}
	\caption{The angular momentum deposition function $\partial \Fdep/\partial R$ for our 3D (top) and 2D models (bottom) as a function of the cooling timescale defined through Eqs.~\eqref{eq:beta-3D}~\&~\eqref{eq:beta-2D} for 3D and 2D, respectively. The quantity approaches zero within $\xshock$ (dashed vertical lines) for very short or very long cooling timescales (blue and red lines, respectively), but peaks around $\beta\sim1$ (gray dots).}
	\label{fig:fdep-beta}
\end{figure}
\begin{figure}
	\centering
	\includegraphics[width=\columnwidth]{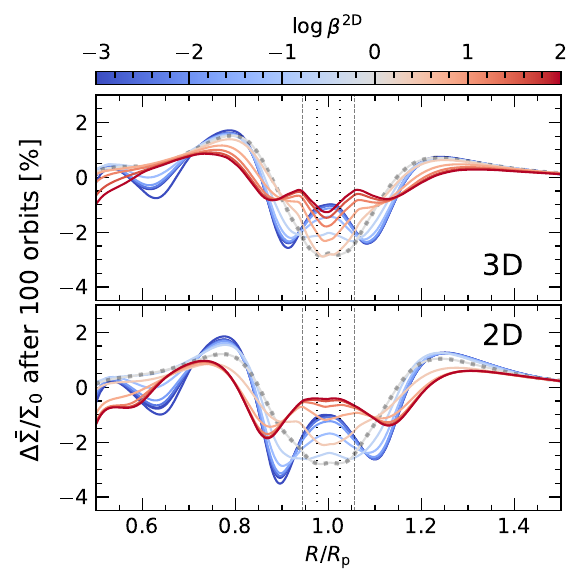}
	\caption{Estimated gap structures after 100 planetary orbits similar to Fig.~\ref{fig:sdot-eos} as a function of $\beta$. The gap structure transitions from double- to single-trough as $\beta$ increases towards $\beta\sim1$, and then back to a double-trough gap for $\beta\gg1$. A secondary gap is also visible for $R\sim0.6\,\Rp$ for $\beta\ll1$ and $\beta\gg1$, induced by the planet's secondary inner spiral arm.}
	\label{fig:sdot-beta}
\end{figure}
We now carry out a suite of 2D and 3D models where we vary the cooling timescale $\beta^\mathrm{2D}$ while keeping the optical depth constant at $\tau=10$, similar to \citet{ziampras-etal-2023a}. We then measure both the angular momentum deposition function $\partial \Fdep/\partial R$ and the expected gap structure $\Delta\bar{\Sigma}/\Sigma_0$ (via the surface density evolution rate $\partial \Sigma/\partial t$) as a function of $\beta$.

Our results for our fully radiative models in both 2D and 3D are shown in Figs.~\ref{fig:fdep-beta}~and~\ref{fig:sdot-beta}. Specifically, in Fig.~\ref{fig:fdep-beta} we see how very little angular momentum is deposited within $|R-\Rp|<\xshock$ for very short or very long cooling timescales (blue or red limits, respectively), but the angular momentum deposition increases as we approach $\beta\sim1$, highlighted with gray dots. Similar trends can be seen in Fig.~\ref{fig:sdot-beta}, where the gap structure transitions from double- to single-trough as $\beta$ increases from $\beta\ll1$ to $\beta\sim1$, and then back to a double-trough gap for $\beta\gg1$. A secondary gap is also visible for $R\sim0.6\,\Rp$ for $\beta\ll1$ and $\beta\gg1$, induced by the planet's secondary spiral arm in the inner disk. This spiral is strongly suppressed and cannot deposit a substantial amount of angular momentum for $\beta\sim1$, resulting in a lack of additional gaps by a single planet, in line with the findings of \citet{ziampras-etal-2020b} and \citet{miranda-rafikov-2020a}.

Finally, we would like to quantify the contribution of radiative damping to gap opening as a function of $\beta$. To do this, we work under the assumption that the only mechanism of angular momentum deposition that contributes to gap opening is the one due to planetary spirals, and that it can be decomposed into two components: one due to shocks beyond $\xshock$, and one due to radiative damping within $\xshock$. Since $\partial\Fdep/\partial R$ switches signs about $\Rp$ (and would therefore sum to zero over the gap-opening region), we choose $\partial\Sigma/\partial t$ as a better indicator of gap opening near the planet. With these assumptions, we define the gap opening efficiency factor $G$ as 
\begin{equation}
	\label{eq:Gdep}
	G = \frac{1}{\Omega_\mathrm{p}\Rp}\int\limits_{\Rp-\xshock}^{\Rp+\xshock} \frac{1}{\Sigma}\frac{\partial \Sigma}{\partial t}\,\mathrm{d}R.
\end{equation}
In other words, $G$ reflects the mass flux exiting the gap region within the shock distance $\xshock$ from the planet. Since we are interested in relative changes in $G$ due to cooling, we normalize it to the value for our adiabatic 2D model, $G_\mathrm{adb}^\mathrm{2D}$ when plotting.

\begin{figure}
	\centering
	\includegraphics[width=\columnwidth]{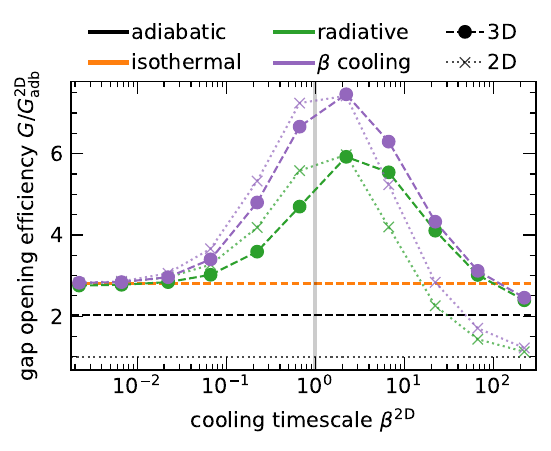}
	\caption{The gap opening efficiency factor $G$, normalized to the adiabatic 2D model value $G_\mathrm{adb}^\mathrm{2D}$, as a function of the cooling timescale $\beta$ for our fully radiative models (green) and those with local, $\beta$ cooling (purple), finding a good agreement between 2D and 3D models. Local $\beta$-cooling overestimates $G$ compared to fully radiative models.}
	\label{fig:G}
\end{figure}

We then present our results in Fig.~\ref{fig:G}, where we show that $G$ changes substantially as a function of $\beta$, peaking for $\beta\sim1$ and recovering the isothermal and adiabatic values for $\beta\ll1$ and $\beta\gg1$, respectively. These findings are consistent with those of \citet{ziampras-etal-2023a}, even though a different metric was used to quantify the gap opening efficiency. Compared to that study, here we can obtain a more appropriate measurement of the gap opening efficiency in the vicinity of the planet rather than the entire disk.

From Fig.~\ref{fig:G} we can also see that the 2D and 3D models agree very well in terms of the shape and amplitude of $G$ across different $\beta$, albeit with a slight offset in the value of $\beta$ at which the peak occurs. More importantly, however, we find that models with local, $\beta$ cooling (purple lines) consistently overestimate the contribution of radiative damping to gap opening compared to fully radiative models (green lines). We interpret this as a result of the local cooling approach not properly capturing the effects of radiative diffusion, but rather assuming that the cooling is entirely localized around the spiral arms.
\subsection{Comparison between 2D and 3D models}
\label{sub:results-2d-3d}

When discussing our findings in previous sections, we noted a very good agreement between our 2D and 3D models in terms of the metrics we used to quantify gap opening. Both the angular momentum deposition function $\partial \Fdep/\partial R$ and the expected surface density evolution rate $\partial \Sigma/\partial t$ show both qualitatively and quantitatively similar radial behavior in Figs.~\ref{fig:fdep-eos}~\&~\ref{fig:sdot-eos}, and the gap opening efficiency factor $G(\beta)$ in Fig.~\ref{fig:G} also behaves very similarly between the two frameworks. This result paints an optimistic picture for the use of 2D radiative models in the context of planet-driven gap opening \citep[see also][for a complementary focus on isothermal models]{fung-chiang-2016,cordwell-etal-2025}. They can be run with a much lower computational cost than their 3D counterparts while recovering the same gap structure, its dependence on the cooling timescale, and even the overestimation of $G$ in models with $\beta$ cooling compared to fully radiative ones.

\begin{figure}
	\centering
	\includegraphics[width=\columnwidth]{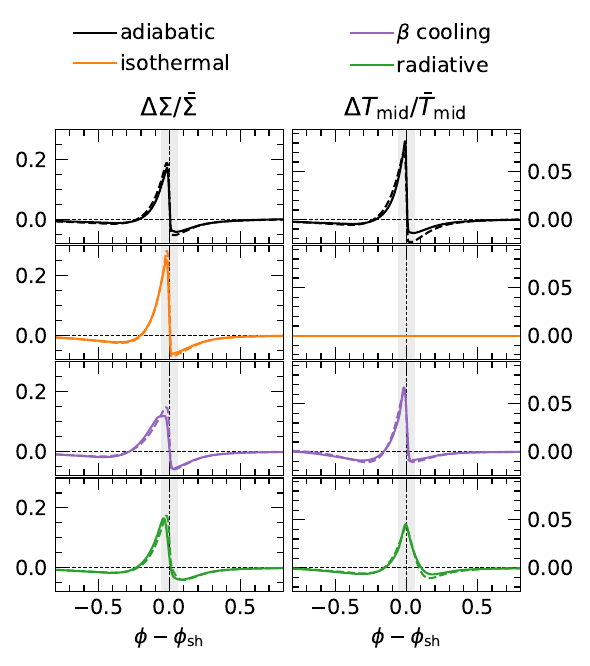}
	\caption{Azimuthal structure of the perturbed surface density (left) and midplane temperature (right) across the shock fronts at $R=1.15\,\Rp$ in our fiducial models for all cooling treatments. The 2D and 3D models (dashed and solid lines, respectively) agree very well with each other.}
	\label{fig:shock-fronts}
\end{figure}

Another noteworthy similarity can be found in the amplitude and shape of the surface density and midplane temperature deviations across the shock fronts themselves, as shown in Fig.~\ref{fig:shock-fronts} for our fiducial set with $\beta^\mathrm{2D}\approx2$ and at $R=1.15\Rp$, where $\partial\Fdep/\partial R$ due to shocks is maximized (see Fig.~\ref{fig:fdep-eos}). In this figure, we find that the 2D and 3D models agree remarkably well across all equations of state. This is particularly interesting since 3D shocks are expected to be weaker than their 2D counterparts \citep{lyra-etal-2016}, but this is not straightforward to infer from this figure.

\begin{figure}
	\centering
	\includegraphics[width=\columnwidth]{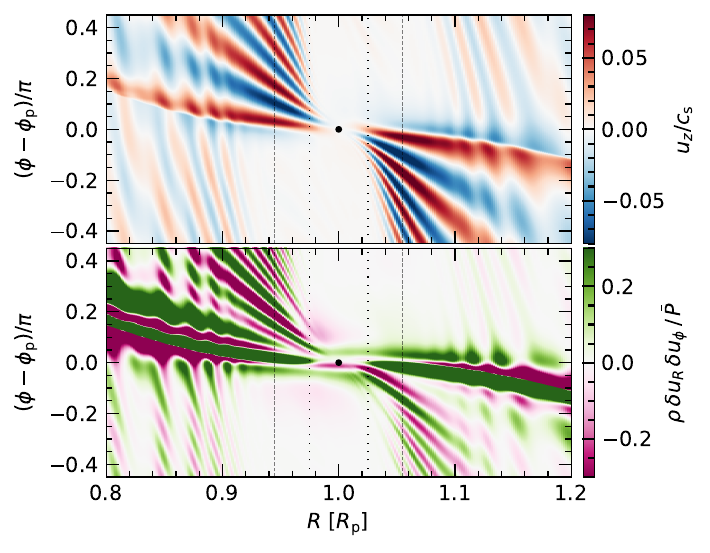}
	\caption{Heatmaps of the vertical velocity (top) and the density-weighted correlation between the radial and azimuthal velocity components (bottom) for our adiabatic 3D model at $z=2H$. Buoyancy modes are clearly visible as diagonal stripes in the top panel, while the bottom panel shows a nonzero angular momentum flux around the planet (black dot).}
	\label{fig:buoyancy}
\end{figure}

Nevertheless, some differences between the two frameworks are expected. The clearest deviation between 2D and 3D is found in the angular momentum deposition profiles for our adiabatic models in Fig.~\ref{fig:fdep-eos}, where the 3D model shows a noticeably lower $\partial \Fdep/\partial R$ beyond $\xshock$ but also a nonzero contribution within the planet's corotating region. The latter is most likely due to the disk's buoyancy response \citep{zhu-etal-2012}, which can drive substantial radial and vertical motion in the region between the horseshoe edge and the shock distance \citep{mcnally-etal-2020,ziampras-etal-2023b} and is therefore not filtered out by our smoothing procedure. In Fig.~\ref{fig:buoyancy} we show heatmaps of the vertical velocity and the density-weighted correlations between the radial and azimuthal velocity components for our adiabatic 3D model at $z=2H$, highlighting both the presence of buoyancy modes and a nonzero angular momentum flux in the vicinity of the planet. In contrast, no such features can be captured in a 2D model.

We note that this is different from the vortensity striping reported in \citet{cordwell-etal-2025}, which referred to a transient process during the horseshoe mixing phase in their isothermal model rather than the active vortensity evolution due to buoyancy modes in our adiabatic run. In fact, it is unclear whether the contribution to $\partial \Fdep/\partial R$ due to buoyancy modes will persist as the gap opening process develops. For this reason, the third, gap-like feature at $R=\Rp$ in Figs.~\ref{fig:sdot-eos}~\&~\ref{fig:fdep-beta} should be interpreted with caution. Clarifying this could be the subject of future work.

Finally, while the agreement between 2D and 3D models is excellent in several aspects in terms of the gap opening efficiency factor $G$ in Fig.~\ref{fig:G}, we note that the 3D models show a peak at $\beta\approx2$ rather than $\beta\approx1$ as in the 2D models. A difference between the two frameworks is expected (at least in the case of $\beta$ cooling), as in 3D models the disk cools over the entire vertical column at each point along the disk plane, whereas 2D models attempt to capture this process in a vertically integrated fashion. Nevertheless, an agreement within a factor of two in terms of the value of $\beta$ at which gap opening is most efficient is still very good, considering the significantly higher computational cost of 3D models and the excellent agreement between 2D and 3D regarding the behavior of $G$ as a function of $\beta$ in both amplitude and shape.

\section{Discussion}
\label{sec:discussion}

In this section, we discuss the implications of our findings in the context of planet-driven gap opening. We also address the limitations of 2D models in capturing the full complexity of the thermodynamic aspects of planet--disk interaction.

\subsection{Cooling in parameter studies of planet--disk interaction}
\label{sub:discussion-cooling}

Modeling radiative processes in a satisfactory manner has been historically notoriously difficult in the context of protoplanetary disks. For one, the cooling timescale is highly sensitive to both gas and dust parameters such as the opacity, the dust-to-gas ratio, the gas surface density, and the temperature. Given the massive uncertainties on these parameters, and the sensitivity of the gap opening process to the cooling timescale, it may even seem counterproductive to include radiative effects in models of planet--disk interaction. At the same time, the high computational cost of radiative simulations makes them impractical for parameter studies, especially in 3D.

However, several groundbreaking advancements have been made in recent years to both constrain disk parameters and develop more efficient numerical methods for radiative transfer. On the observational side, high-resolution ALMA observations have provided stringent constraints on the disk temperature structure in 3D \citep[e.g.,][]{law-etal-2021,galloway-sprietsma-etal-2025}, as well as on the overall disk masses from which surface density profiles can be inferred \citep[e.g.,][]{zhang-etal-2025,trapman-etal-2025,longarini-etal-2025}. Multi-wavelength observations of the same systems have also been utilized to infer dust properties such as the maximum grain size and dust size distribution \citep[e.g.,][]{doi-kataoka-2023,doi-etal-2024}. These constraints, in combination with improved models of dust evolution \citep[e.g.,][]{birnstiel-etal-2012,drazkowska-etal-2019,stammler-birnstiel-2022}, dust opacities \citep[e.g.,][]{woitke-etal-2016,birnstiel-etal-2018,dominik-etal-2021}, and radiative transfer \citep[e.g.,][]{ercolano-etal-2003,dullemond-etal-2012}, can be used to infer the properties of dust grains and constrain disk properties to the extent that a treatment of radiative cooling can be both motivated and yield realistic results. Finally, on the computational side, the rising usage of GPUs as tools for accelerating hydrodynamical simulations \citep[e.g.,][]{benitez-llambay-etal-2016,lesur-etal-2023,stone-etal-2024} in combination with highly optimized linear algebra libraries that can be used for the express purpose of solving stiff operators such as the source terms in FLD \citep[see Eq.~\eqref{eq:Qrad}; e.g.,][]{petsc-web-page,trilinos-website}, have made it feasible to carry out large-scale radiative hydrodynamical simulations in 3D.

From the above it becomes clear that while modeling planet--disk interaction with a treatment for radiative transfer is quite involved and challenging, it is becoming increasingly more feasible and necessary to capture the full complexity of the problem. In fact, simple thermodynamic assumptions (e.g., locally isothermal) fail to capture the intricacies of planet--disk interaction in several ways including gap opening \citep[e.g.,][]{miranda-rafikov-2020a,ziampras-etal-2023a}, planet-induced vortex dynamics \citep[e.g.,][]{fung-ono-2021,rometsch-etal-2021}, circumplanetary disk dynamics \citep[e.g.,][]{krapp-etal-2024}, gap edge dynamics \citep[e.g.,][]{muley-etal-2024}, and planet migration \citep[e.g.,][]{lega-etal-2014,ziampras-etal-2024a,ziampras-etal-2024b,ziampras-etal-2025b}.

Nevertheless, our results show that in the context of planet-induced gap opening it is both possible and accurate to use 2D models with a simple yet physically motivated treatment of radiative cooling via Eqs.~\eqref{eq:bmid-bsurf-2D}~\&~\eqref{eq:beta-2D}~or~\eqref{eq:beta-2D-f}. This opens up the possibility of carrying out large parameter studies of planet--disk interaction in 2D, while still capturing the key features of fully radiative models.

\subsection{The role of viscous diffusion in gap opening}
\label{sub:discussion-viscosity}

Viscosity acts as both a damping mechanism for spiral shocks and as a diffusive process across the gap region. As such, the angular momentum deposition function $\partial \Fdep/\partial R$ is expected to be different in disks with significant levels of viscous diffusion. This could be important when comparing 2D to 3D results, as several sources of turbulence that require the inclusion of the vertical direction have been investigated in the context of protoplanetary disks but cannot be inherently captured in 2D models. Such mechanisms include the magnetorotational instability \citep[MRI,][]{balbus-hawley-1991,hawley-balbus-1991}, the vertical shear instability \citep[VSI,][]{nelson-etal-2013}, the streaming instability \citep[SI,][]{youdin-johansen-2007,johansen-youdin-2007}, the spiral wave instability by the planet's spiral shocks \citep[SWI,][]{bae-etal-2016a,bae-etal-2016b}, the convective overstability \citep[COS,][]{klahr-hubbard-2014,lyra-2014}, and the zombie vortex instability \citep[ZVI,][]{marcus-etal-2015,marcus-etal-2016}.
Nevertheless, as \citet{miranda-rafikov-2020b} have shown, the effects of viscosity on the spiral angular momentum flux are negligible for $\alpha\lesssim10^{-2}$ (see Fig.~12 therein) which is generally the case for the aforementioned mechanisms and in realistic disk conditions \citep[see][for further details]{pfeil-klahr-2019,lyra-umurhan-2019,lesur2023}.

It should be noted, however, that while the general shape of the gap is expected to be similar between 2D and 3D models, the detailed structure of the gap edges may differ \citep[see also][]{fung-chiang-2016}. This is particularly important in the context of vortices driven by the Rossby-wave instability \citep[RWI,][]{lovelace-1999,chang-youdin-2024}, as their formation is sensitive to the shape of the gap edges and they can act as a source of turbulent diffusion during their lifetime \citep{lega-etal-2021}. This effect is especially relevant for low-viscosity models such as the inviscid ones presented here, and could be the subject of future work.

\section{Summary}
\label{sec:summary}

We performed high-resolution two- and three-dimensional hydrodynamical simulations featuring embedded gap-opening planets. Our intent was to compare the 2D and 3D frameworks in the context of planet-driven gap opening across different thermodynamical treatments: locally isothermal, adiabatic, local $\beta$ cooling, and full radiation transport including the effects of radiative diffusion. In doing so, we could both examine the accuracy of 2D models with respect to 3D, and identify simplified cooling prescriptions that could reproduce the key features of fully radiative models reasonably well.

We found that the overall angular momentum deposition profiles $\partial\Fdep(R)/\partial R$---which in turn determine the gap opening efficiency and resulting gap shape---were very similar between 2D and 3D models for the same thermodynamical treatment. This is in good agreement with the findings of \citet{cordwell-etal-2025} and \citet{fung-chiang-2016}, who focused on locally isothermal models. Given the profiles of $\partial\Fdep(R)/\partial R$ we could then synthesize the gap structure at its early phase following \citet{cordwell-rafikov-2024}, and demonstrated that the expectation of a ``single trough'' gap shape for $\beta\sim1$ and a ``double trough'' gap shape for locally isothermal and adiabatic models holds, in excellent agreement with previous studies \citep[e.g.,][]{miranda-rafikov-2020a}.

We then investigated the dependence of the gap opening efficiency as a function of the cooling timescale by examining the total angular momentum deposition $\partial\Fdep(R)/\partial R$ in the vicinity of the planet for different cooling prescriptions and disk parameters. Here, we found a continuous transition from double- to single- to double-trough gap shapes as $\beta$ varied from $\ll1$, to $\sim1$, and finally to $\gg1$, in line with the above expectations. We also found our metric for the gap opening efficiency $G(\beta)$ to peak for $\beta\sim1$ as a direct consequence of that.

We found a very good match between 2D and 3D models in terms of the dependence of $G(\beta)$ on the cooling timescale, albeit with a few quantitative differences. In particular, models with local $\beta$ cooling slightly overestimated the overall gap opening efficiency compared to their fully radiative counterparts, and the peak in $G(\beta)$ for 2D models was found to be slightly offset towards shorter $\beta$ by a factor of $\approx2$.

Nevertheless, the very good overall agreement between 2D and 3D models across all thermodynamical prescriptions is encouraging for two reasons. First, it suggests that 2D simulations can be used to accurately capture the gap structure around embedded planets, even when radiative effects are important. This is particularly relevant in the context of modeling the substructures found in observed systems with embedded planets, as 2D models are significantly less computationally expensive than their 3D counterparts. Second, it indicates that the use of a local, ``$\beta$ cooling'' prescription can be a viable alternative to the more complex fully radiative approach when studying gap opening by planets. This makes it much easier to explore and/or constrain the parameter space before executing dedicated---possibly 3D---simulations with full radiation transport.

We believe that our results provide a solid foundation for the use of radiative processes in future studies of planet--disk interaction, and encourage the inclusion of radiation transport, even with simplified prescriptions, in such studies.

\begin{acknowledgements}
This research utilized Queen Mary's Apocrita HPC facility, supported by QMUL Research-IT (http://doi.org/10.5281/zenodo.438045). This work was performed using the DiRAC Data Intensive service at Leicester, operated by the University of Leicester IT Services, which forms part of the STFC DiRAC HPC Facility (www.dirac.ac.uk). The equipment was funded by BEIS capital funding via STFC capital grants ST/K000373/1 and ST/R002363/1 and STFC DiRAC Operations grant ST/R001014/1. DiRAC is part of the National e-Infrastructure. AZ acknowledges funding from the European Union under the European Union's Horizon Europe Research and Innovation Programme 101124282 (EARLYBIRD). AJC is funded by the Royal Society of New Zealand Te Apārangi and the Cambridge Trust through the Cambridge--Rutherford Memorial Scholarship. RRR acknowledges financial support through the STFC grant ST/T00049X/1 and the IAS. RPN is supported by STFC grant ST/X000931/1. Views and opinions expressed are those of the authors only. All plots in this paper were made with the Python library \texttt{matplotlib} \citep{hunter-2007}. Typesetting was expedited with the use of GitHub Copilot, but without the use of AI-generated text.
\end{acknowledgements}

\section*{Data Availability}

Data from our numerical models are available upon reasonable request to the corresponding author.

\bibliographystyle{aa}
\bibliography{refs}


\appendix

\section{Resolution study}
\label{apdx:resolution}

To ensure that our results are appropriately converged with respect to numerical resolution, we carried out a series of runs for our fiducial 3D model with $\beta^\mathrm{2D}\approx2$ where we varied the resolution from 4 to 32 cells per scale height (cps) at the planet's location. The same resolution check for 2D models was carried out in \citet{ziampras-etal-2023a}, where it was found that 32~cps were sufficient for convergence, so we did not repeat it here.

\begin{figure}
	\centering
	\includegraphics[width=\columnwidth]{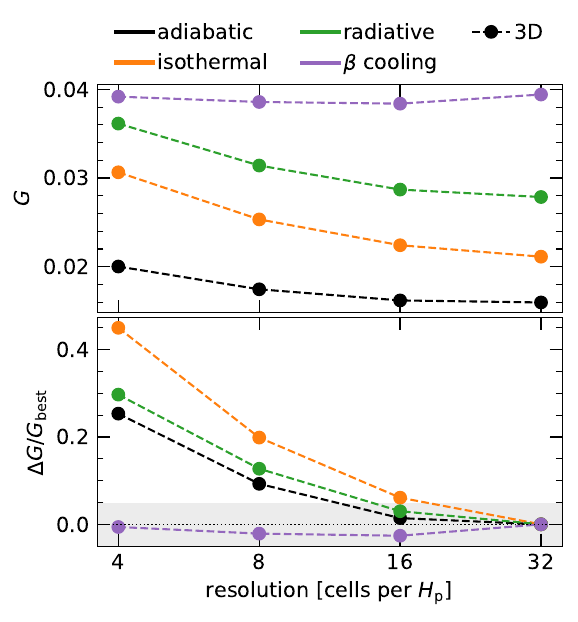}
	\caption{Gap opening efficiency metric $G$ as a function of the number of grid cells per scale height at the planet's location. Top: absolute value of $G$. Bottom: relative difference with respect to the highest-resolution run with 32~cps. We find that $G$ is converged to within $\approx5$\% for 16~cps.}
	\label{fig:resolution}
\end{figure}

Figure~\ref{fig:resolution} shows our gap opening efficiency metric $G$ (see Eq.~\eqref{eq:Gdep}) as a function of the number of grid cells per scale height at the planet's location. We find that $G$ is converged to within $\approx5$\% for 16~cps with respect to the 32~cps run. Given the high computational cost of our fully radiative 3D models, we chose to run our main suite of 3D models with 16~cps, which we consider to be sufficiently converged for our purposes.
	
\section{Derivation of the cooling timescale}
\label{apdx-beta}

Here we aim to provide a heuristic derivation of the dimensionless cooling timescale $\beta$ used in Eq.~\eqref{eq:beta-3D}. Our approach relies on the following key assumptions:
\begin{itemize}
	\item if cooling happens via dust thermal emission, the collisional coupling timescale between gas and dust is far shorter than the cooling timescale \citep[see][for details]{dullemond-etal-2022};
	\item the radiation field $\Erad$ is isotropic and constant in time, and approximately satisfies $\Erad\approx\aR T^4$ (one-temperature approximation within the FLD framework);
	\item cooling due to advection or adiabatic decompression (terms $\vel\nabla e$ and $-\gamma e\nabla\cdot\vel$ in Eq.~\eqref{eq:navier-stokes-3}) is negligible.
\end{itemize}
Equation~\eqref{eq:navier-stokes-3} can then be written as
\begin{equation}
	\label{eq:navier-stokes-3-simple}
	\DP{e}{t} \approx -\nabla\cdot\bm{F} \Rightarrow \rho\cv\D{T}{t} \approx \nabla\cdot\left(\frac{\lambda c}{\kappa\rho}\nabla \aR T^4\right),
\end{equation}
We now consider the two limits where cooling is limited by radiative diffusion due to the medium being opaque to radiation (optically thick,  $\lambda\rightarrow 1/3$) or by the emissivity of the medium (optically thin, $F\rightarrow c\aR T^4$).

For simplicity, we will further assume that quantities other than temperature are constant such that they can be pulled out of the gradient operators. In practice, our results can be corrected with a prefactor of order unity to account for a typical radial and/or temperature dependence of $\rho$ or $\kappa$.

In the optically thick limit, Eq.~\eqref{eq:navier-stokes-3-simple} transforms into a diffusion equation with
\begin{equation}
	\label{eq:navier-stokes-3-diffusion}
	\rho\cv\DP{T}{t} \approx \nabla\cdot\left(\frac{c\aR 4 T^3}{3\kappa\rho}\nabla T\right) \Rightarrow \DP{T}{t} \sim \eta\nabla^2 T,
\end{equation}
where we have used that $\aR c = 4\sigmaSB$ and the definition of the radiative diffusion coefficient $\eta$ from Eq.~\eqref{eq:beta-3D}. We can now estimate the diffusion timescale from the above equation as $t_\text{thick} = \beta_\text{thick}\OmegaK^{-1} = l_\text{diff}^2/\eta$, where $l_\text{diff}$ is a characteristic diffusion lengthscale. Assuming that $l_\text{diff}\sim H$ \citep{flock-etal-2017a,ziampras-etal-2023a}, we arrive at
\begin{equation}
	\beta_\text{thick} = \frac{\OmegaK}{\eta} H^2.
\end{equation}

In the optically thin limit, Eq.~\eqref{eq:navier-stokes-3-simple} instead becomes
\begin{equation}
	\label{eq:navier-stokes-3-advection}
	\rho\cv\DP{T}{t} \approx \nabla\cdot\left(c\aR T^4\right) \Rightarrow \DP{T}{t} \approx \bm{V}\cdot\nabla T,\quad V = \frac{16\sigmaSB T^3}{\rho\cv}
\end{equation}
which corresponds to an advection of freely streaming radiation at a characteristic velocity $\bm{V}$. Here, the characteristic lengthscale is the photon mean free path $\lrad$ (see Eq.~\eqref{eq:beta-3D}), and we can estimate the advection timescale as $t_\text{thin} = \beta_\text{thin}\OmegaK^{-1} = \lrad/V$. We then arrive at
\begin{equation}
	\beta_\text{thin} = \frac{\OmegaK \rho\cv \lrad}{16\sigmaSB T^3} = \frac{\OmegaK}{\eta}\frac{\kappaP}{\kappaR}\frac{\lrad^2}{3}.
\end{equation}

Realistically, cooling will be primarily limited by the slowest of the two channels. We can therefore estimate the total cooling timescale as
\begin{equation}
	\beta_\text{3D} = \beta_\text{thick} + \beta_\text{thin} = \frac{\OmegaK}{\eta}\left(H^2 + \frac{\kappaP}{\kappaR}\frac{\lrad^2}{3}\right),
\end{equation}
which matches Eq.~\eqref{eq:beta-3D} in the main text \citep[a similar, although slightly different in its final form, result was obtained in][]{miranda-rafikov-2020b}. Alternatively, one can write that $\beta_\text{3D} = \max\left(\beta_\text{thick}, \beta_\text{thin}\right)$ \citep[e.g.,][]{pfeil-klahr-2021} which behaves similarly but is not differentiable at the transition between the two regimes.

\end{document}